%% file: Tam2013_arxiv.tex
\documentclass[a4,twoside,10pt]{article}
\usepackage{amsmath,amssymb}
\usepackage{tam13}
\usepackage{xspace}
\usepackage{bm}
\usepackage[font=footnotesize,labelfont=rm]{caption}
\usepackage{hyperref}
\input{config}
\title{The Discovery of the Higgs Boson with the CMS Detector and its Implications for Supersymmetry and Cosmology}
\shorttitle{Discovery of the Higgs Boson}
\authors{W. de Boer$^1$\email{wim.de.boer@kit.edu},
for the CMS Collaboration}
\shortauthors{W. de Boer}
\affiliations{
$^1$Institut f\"ur Experimentelle Kernphysik, Karlsruhe Institute of Technology, P.O. Box 6980, 76128 Karlsruhe, Germany
}

\abstract{
The discovery of the long awaited Higgs boson is described using data from the CMS detector at the LHC. In the SM the masses of fermions and the heavy gauge bosons are generated by the interactions with the Higgs field, so all couplings are related to the observed masses. Indeed,  all observed couplings are consistent with the predictions from the Higgs mechanism, both to vector bosons and fermions implying that masses are indeed consistent of being generated by the interactions with the Higgs field. However, on a cosmological scale the mass of the universe seems not to be related to the Higgs field: the baryonic mass originates from the binding energy of the quarks inside the nuclei and dark matter is not even predicted in the SM, so the origin of its mass is unknown. The dominant energy component in the universe, the dark energy, yields an accelerated expansion of the universe, so its repulsive gravity most likely originates from a kind of vacuum energy. The Higgs field would be the prime candidate for this, if the energy density would not be many orders of magnitude too high, as will be calculated. The Higgs mass is found to be 125.7$\pm$0.3(stat.)$\pm$0.3(syst.) GeV, which is below 130 GeV, i.e. in the range predicted by supersymmetry. This  may be the strongest hint for supersymmetry in spite of the fact that the predicted supersymmetric particles  have not been discovered so far.
}
\begin{document}
\maketitle
\newpage
\tableofcontents
\section{Introduction}
\label{Introduction}
The discovery of a 126 GeV Higgs boson \cite{Aad:2012tfa,Chatrchyan:2012ufa}  is  exciting for several reasons:
i) Its branching ratios to the electroweak gauge bosons  are in agreement with the predictions from the Higgs mechanism, thus providing evidence that masses of gauge bosons are generated by interactions. This  eliminates
the need for explicit mass terms in the Lagrange density of the Standard Model (SM), which would violate local gauge invariance and therefore ruins  the basis of the SM. For a review of the Higgs me\-chanism in the SM, see Ref. \cite{Djouadi:2005gi}.
ii) In the SM electroweak symmetry breaking  is introduced ad hoc and the Higgs boson mass is not predicted, so it could have  any value between the electroweak scale and the TeV scale.  However, in the supersymmetric extension of the SM (SUSY) EWSB {\it is} predicted (by radiative corrections) and the mass of the lightest Higgs boson is predicted to be below 130 GeV, as observed.
This  may be the strongest hint for SUSY in spite of the fact that the predicted supersymmetric particles  have not been discovered so far. For a review of the Higgs sector in  SUSY, see Ref. \cite{Djouadi:2005gj}.
Additional hints for SUSY are the unification of gauge and Yukawa couplings at a large scale, 
the GUT scale, as expected in Grand Unified Theories, the absence of quadratic divergencies to the Higgs 
boson mass, and the prediction of a dark matter particle with a correct relic density, 
see reviews, e.g. \cite{Haber:1984rc,deBoer:1994dg,Jungman:1995df,Martin:1997ns,Kazakov:2010qn}.

Details of the CMS Higgs boson discovery can be found in Ref. \cite{Chatrchyan:2013lba} and an update 
of the combined results with the full luminosity and other updated results can be found on the public CMS webpages \cite{CMS:yva}. 
The results reported here are partly based on these not yet published, but updated data.

The paper is structured as follows: we start with a short introduction to the Higgs mechanism and 
its expected properties following from this mechanism. In the following section we discuss 
the discovery of the Higgs boson and compare its properties with the SM expectation. 
We proceed by discussing its cosmological implications, especially the fact that the dark energy 
predicted by the Higgs field is many orders of magnitude larger than observed, even in the supersymmetric 
extension of the SM. We show also that the Higgs particle has actually little to do with 
the mass of the universe. 
We finish by discussing the implications of the observed Higgs boson mass for supersymmetry
and its possible role in cosmology.
\section{The SM and its Higgs mechanism}
\subsection{Introduction}
The SM does not allow mass terms for the weak gauge bosons, but this problem 
can be solved by assuming that  masses are generated
dynamically through the interaction with a scalar field, omnipresent in the vacuum.
In this case the SM stays renormalizable, as was proven by 't Hooft and Veltman \cite{'tHooft:1972fi}, who were awarded the Nobel prize for this in 1999.

The use of a field in the vacuum to generate masses was  first proposed by
Schwinger in 1962 \cite{Schwinger:1962tp} and it was applied in 1963 by Anderson \cite{Anderson:1963pc}  to the non-relativistic case of superconductivity, in which the rotational symmetry of the system is broken, when entering the superconducting phase. The formulation of a consistent field theory possessing a broken symmetry was hampered by  Goldstone's  theorem, which predicted massless, scalar bosons \cite{Goldstone:1961eq,Goldstone:1962es,Bludman:1963zza} after symmetry breaking.
However, such massless, scalar particles had not been observed. 
Anderson argued that the matter spectrum before including the Yang-Mills interaction contains these massless (Goldstone) bosons, but that the example of superconductivity illustrates
that the physical spectrum need not. In superconductivity the Cooper pairs form a scalar field and  the Meissner effect  can be viewed as generating a mass for the photon.  

The possibility of a relativistic "Higgs mechanism" was proposed by Higgs  in 1964 \cite{Higgs:1964ia} for the abelian QED-like example and later extended
to a non-abelian SU(3) example   \cite{Higgs:1964pj}.
He writes: {\it "The purpose of the present note is to report that,
as a consequence of the coupling  between gauge bosons and the scalar field, the spin-one
quanta of some of the gauge fields acquire mass;
the longitudinal degrees of freedom of these particles
(which would be absent if their mass were
zero) go over into the Goldstone bosons when the
coupling tends to zero. This phenomenon is just
the relativistic analog of the plasmon phenomenon
to which Anderson' has drawn attention:
that the scalar zero-mass excitations of a superconducting
neutral Fermi gas become longitudinal
plasmon modes of finite mass when the gas
is charged."}
So he establishes the relation between the Goldstone massless bosons, which disappear after the gauge bosons get
their mass. Englert and  Brout had shown just before  that in non-Abelian gauge theories
spontaneous symmetry breaking could give masses to gauge bosons \cite{Englert:1964et}. However, nothing was said about the disappearance of the problematic, massless Goldstone bosons.
Later,  Guralnik, Hagen and Kibble showed, that in a specific model,
spontaneous symmetry breaking can lead to massive gauge bosons
and the  absence of massless bosons in a spontaneously broken symmetry is a consequence of the inapplicability
of Goldstone's theorem rather than a contradiction of it \cite{Guralnik:1964eu}.

Generating mass means the particle is slowed down in the medium, as it follows 
directly from the relativistic energy-momentum relation: $E^2=p^2+m^2$ (using natural units in the whole paper, i.e. c=1).
If the mass is zero, the particle must have a velocity $\beta=v/c=p/E=1$, but if
the mass term is positive, then necessarily $\beta=p/E=\sqrt{E^2-m^2}/E \le 1$.
For elementary particles the medium can be the   vacuum, if it is filled with a field slowing down the particle,
in this case the Higgs field, as we know now.  Later on, we will see that  the weak gauge bosons get their mass from the kinetic terms in the Lagrangian, which leads to a mass term from the interaction with the scalar field.  A pictorial analogy would be the kinetic energy of a swimmer, which is reduced  after diving into the water, thus creating mass by  the interaction with the "water" field, the $H_2O$ molecules being its quanta. 

If the vacuum slows down the particles, it must be filled with a field with a nonzero vacuum expectation value (vev), which in case of a complex scalar field means that the phases are not randomly oriented, but must have a fixed value, so the field can be represented by $\Phi=v~exp(i\zeta)$ with the phase $\zeta$ fixed and
$v$ being the vev.  
The same amplitude $v$ of the complex field $\Phi$ is reached
for  arbitrary values of the phase $\zeta$, 
so there exists an infinity
of different, but equivalent ground states.
 This degeneracy of the ground state
takes on a special significance in a quantum field theory,
because the vacuum is required to be unique, so the phase
can't be arbitrarily at each point in space-time.
Once a particular value of the phase is chosen, it has to remain
the same everywhere, i.e. it can't change
locally.
A scalar field with a nonzero vev
therefore breaks the rotational symmetry of the ground state spontaneously. In case of the $SU(2)\otimes U(1)$
symmetry it is called spontaneous electroweak symmetry breaking (EWSB). In the SM this EWSB has to be introduced ad hoc, but in supersymmetry it is induced by radiative corrections, as will be discussed later.
%
\begin{figure}[t]
\centering
\includegraphics[width=0.4\textwidth,height=0.25\textwidth]{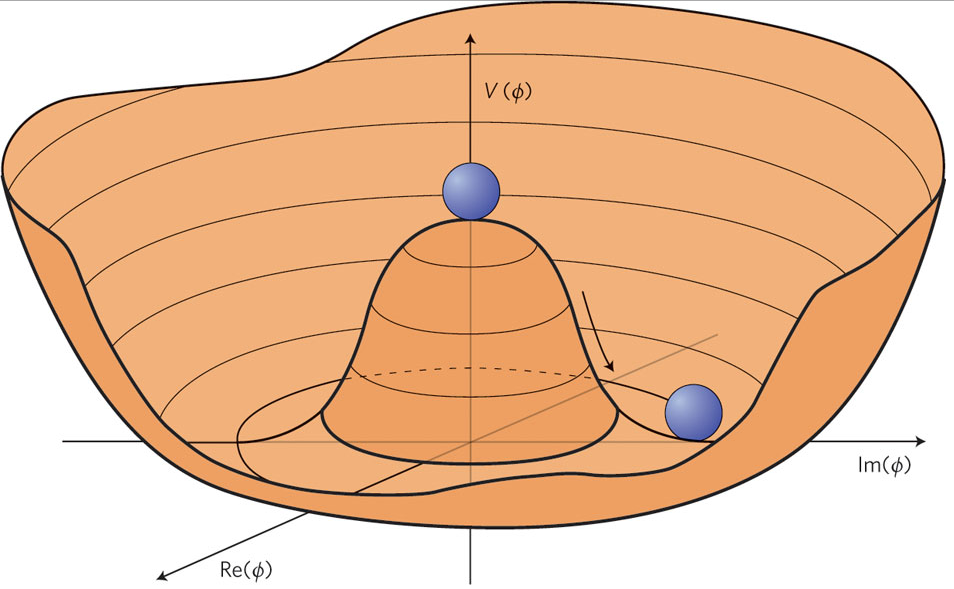}
\caption{ Shape of the Higgs potential for  $\mu^2<0$.}
\label{f1}
\end{figure}
For the breaking of the electroweak symmetry by the Higgs field one considers
a potential ana\-logous to the one proposed by
Ginzburg and Landau for the breaking of the rotational symmetry in superconductivity \cite{Ginzburg:1950sr} \footnote{The Nobel Prize in Physics 2003 was awarded jointly to  Abrikosov,  Ginzburg and  Leggett "for pioneering contributions to the theory of superconductors and superfluids".}:
\bq V(\Phi)=\mu^2\; \Phi^{\dagger} \Phi + \lambda
(\Phi^{\dagger}\Phi)^2, \label{potential}\eq
where $\mu^2$ and  $\lambda$ are constants.
The potential has a parabolic shape, if $\mu^2 > 0$,
but takes the shape of a Mexican hat for $\mu^2<0$,
 as pictured in Fig. \ref{f1}. In the latter case the field free vacuum,
i.e. $\Phi =0 $, corresponds to a local maximum, thus forming
an unstable equilibrium. 
The parameter $\mu^2$ acts like the critical temperature  $T_c$ in superconductivity:
above $T_c$ the electrons are free particles, so  their
phases can be rotated arbitrarily at all points in space,
but below $T_c$ the  rotational freedom is lost,
because the electrons form a coherent system, in which all
phases are locked to a certain value. This corresponds to
a single point in the minimum of the Mexican hat,
 which represents a vacuum with a nonzero vev
 and a well defined phase,
thus defining a unique vacuum.
The coherent system can still be rotated as a whole, so it is
invariant under global but not under local rotations.
\subsection{The Higgs mechanism}
 Detailed up-to-date reviews for the Higgs sector in the SM and supersymmetry
are given  in Refs. \cite{Djouadi:2005gi} and  \cite{Djouadi:2005gj}, respectively.
Here we summarize only the salient features
in order to compare the data with the predictions.
The SM has one massless gauge boson - the photon -, while
the $W$ and $Z$ bosons must be massive. So the Higgs field must couple to the weak
gauge bosons and not to the photon, which implies that it must have weak isospin
and no electric charge. In order not to have a preferred direction in the vacuum of space-time, it should
have no spin, i.e. it should be a scalar ($J^P=0^+$ state) or pseudo-scalar ($J^P=0^-$ state).
This can be achieved by choosing $\Phi$ to be a scalar complex
 $SU(2)$ isospin doublet with definite hypercharge ($Y_W=1$):
\bq \Phi(x)=\frac{1}{\sqrt{2}}\left(\ba{c}\phi_1^+(x)+i\phi^+_2(x)\\
\phi^0_1(x)+i\phi^0_2(x)\ea \right). \label{phix} \eq
The interactions of the Higgs field
with other particles can be obtained from the
Lagrangian for a scalar field:
\bq \LL_H=(D_\mu \Phi)^{\dag} (D^\mu \Phi) -V(\Phi). \label{lhig}\eq
The first term is the usual kinetic energy term for a scalar
particle, for which the Euler-Lagrange equations
lead to the Klein-Gordon equation of motion.
The requirement that  the ground state has to have a nonzero vev and be electrically neutral 
implies that $\Phi$ has to be of the form $\frac{1}{\sqrt{2}}\left(\ba{c}0\\v\ea\right)$.
The quantum fluctuations of the field around the ground state
  can be parametrised as follows, if we include an arbitrary
$SU(2)$ phase factor:
\bq \Phi = e^{i{\bm \zeta}(x)\cdot{\bm\tau}}\frac{1}{\sqrt{2}}\left(\ba{c}0\\v+h(x)\ea\right).\eq
The (real) fields $\bm \zeta(x)$ are  excitations of the field
{\it along} the potential minimum, i.e. changing the phase. They correspond to
the {\it massless Goldstone bosons} of a global symmetry, in this
case three for the three rotations of the $SU(2)$ group.
However, in a {\it local} gauge theory these massless
bosons can be eliminated by a local SU(2) rotation:
\bq \Phi^\prime=e^{-i{\bm \zeta}(x)\cdot{\bm \tau}}\Phi(x)=
\frac{1}{\sqrt{2}}\left(\ba{c}0\\v+h(x)\ea\right).
\label{hmin}\eq
\begin{figure}[t]
\centering
\includegraphics[width=0.6\textwidth]{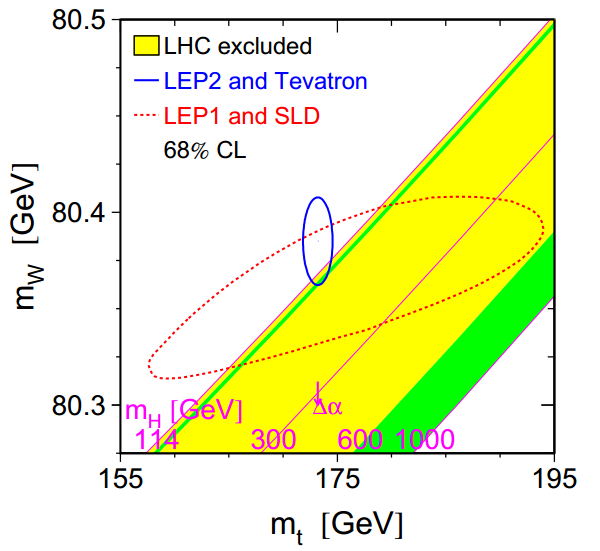}
\caption{ Consistency of gauge boson mass, Higgs boson mass and top mass. From Ref. \cite{LEP-2}.}
\label{f2}
\end{figure}
\begin{figure}[]
\centering
\includegraphics[width=0.7\textwidth]{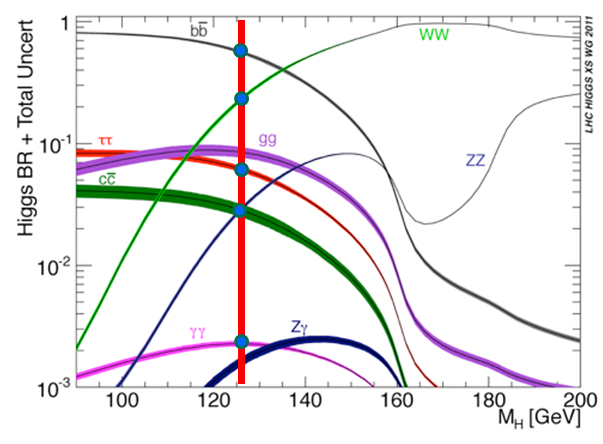}
\caption{Branching fractions of a SM Higgs boson as function of the Higgs mass.  The dots on the vertical line at a Higgs mass of 126 GeV indicate the channels observed by CMS.
}
\label{f3}
\end{figure}
\begin{figure}[t]
\centering
\includegraphics[width=0.6\textwidth]{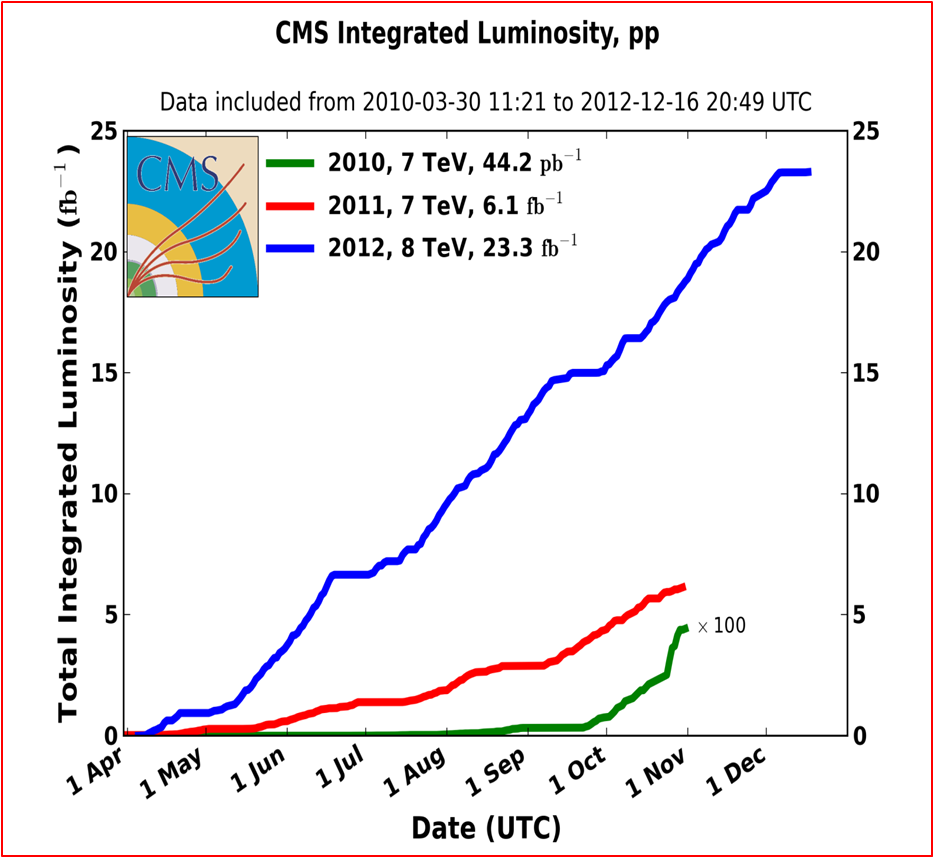}
\caption{ The CMS integrated luminosity as function of time.
}
\label{f4}
\end{figure}
Consequently, the fields $\bm \zeta$ have no physical significance.
Only the real field $h(x)$ can be interpreted as a real
(Higgs) particle. It represents excitations of the field independent of the phase,
 like the ball rolling down in Fig. \ref{f1} for a given phase and oscillating around the minimum,
 which represents one degree of freedom.
The original field $ \Phi$ with four degrees of freedom
has lost three degrees of freedom, which are recovered
as the longitudinal polarisations of the three heavy gauge
bosons. In pictorial language:  the heavy gauge bosons
have eaten the Goldstone bosons and grown fat.

%
\begin{figure}[t]
\centering
\includegraphics[width=0.8\textwidth,height=0.5\textwidth]{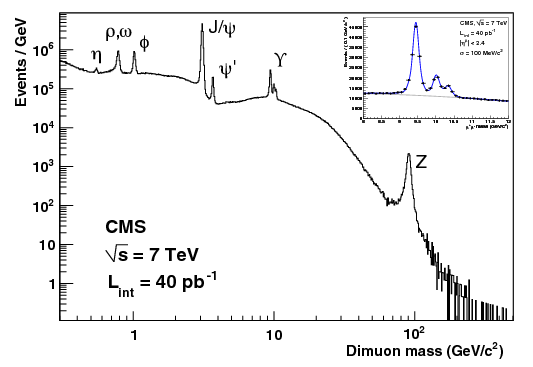}
\includegraphics[width=0.8\textwidth,height=0.5\textwidth]{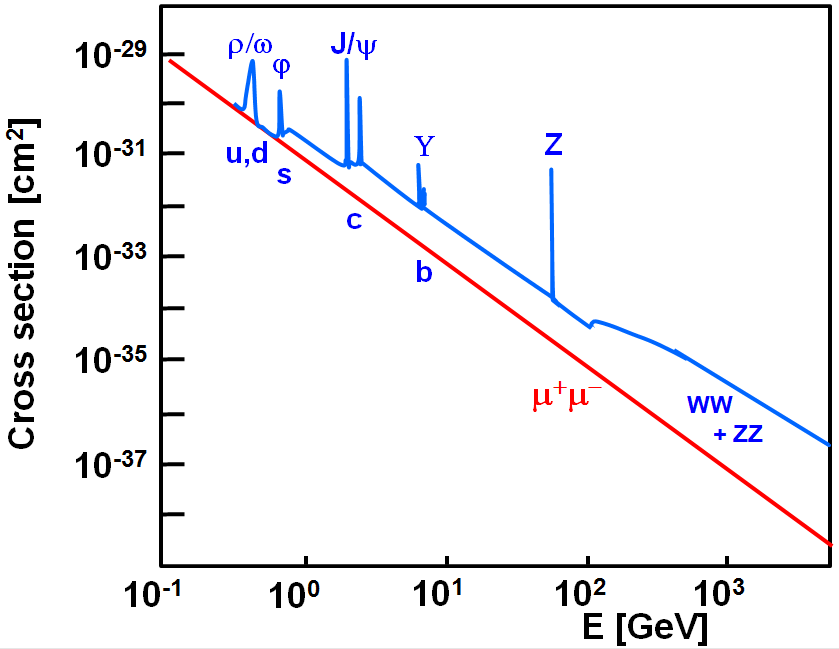}
\caption{Top: Muon pair invariant mass spectrum observed with the CMS detector 
showing the well known resonances. The insert is a zoom around the $\Upsilon$ region. From \cite{Chatrchyan:2012xi}. 
Bottom: the same resonances have been observed at various $e^+e^-$ colliders with increasing energies
over a period of 24 years. The spectrum at the LHC can be obtained within 24 hours with the present luminosity.
}
\label{f5}
\end{figure}
The kinetic part of Eq. \ref{lhig}  gives rise to quadratic terms in the wave function of the vector bosons,
i.e. one obtains mass terms for them from the interaction with the scalar field, the essence of the Higgs mechanism.
The mass terms of the physical fields become:
\bq M_W^2=\frac{1}{4} g^2v^2  ~~~~~
 M_Z^2=\frac{g^{\prime\; 2}+g^2}{4} v^2,
\label{emzz} \eq
where $g$ and $g^\prime$ are the gauge couplings of the SU(2) and U(1) groups of the weak and electromagnetic interactions, respectively.
Given the known gauge boson masses and gauge couplings the vev of the Higgs field can be calculated to be
\bq v \approx 246~ {\rm GeV}. \label{v174}\eq
The value of $M_W$ can  be related to the precisely measured
muon decay constant $G_\mu=1.16639(2)\cdot 10^{-5}$ GeV$^{-2}$.
If calculated in the SM, one finds:
$\mz$=88 GeV.
However, these calculations are only at tree level. Radiative
corrections  to the W and Z masses  involving top and Higgs loops depend on the  top  and Higgs mass.
If the radiative corrections are included, all masses become consistent, as shown in Fig. \ref{f2}  \cite{LEP-2}.
Also the fermions carry weak isospin, so they can interact with the Higgs field, albeit
not necessarily with the gauge coupling constant.
However, 
the Lagrangian for the interaction of the leptons with the Higgs field
has exactly the form of a mass term of a fermion:
\bq \LL_{H-L}=-g_f^e\left[{  \overline{L}{\bm \Phi}
    e_R+\overline{e}_R{\bm \Phi}^{\dag}L}\right]. \eq
The Yukawa coupling constant $g_f^e$ is a free parameter,
 which has to be adjusted
such that \bq m_e=\frac{g_f^e v}{\sqrt{2}}\label{fermionmass}.\eq
Thus  the  Yukawa coupling $g_f$ is proportional to the
mass of the particle  and consequently the coupling of the
Higgs field to fermions is proportional to the mass of the fermion,
a prediction of utmost importance for the experimental search
for the Higgs boson:  the Higgs bosons 
will decay predominantly into b quarks,
if the heavier t quark or W  and Z bosons
are kinematically not allowed.

One may argue that one does not gain anything by rewriting the arbitrary fermion mass in terms of an arbitrary Yukawa coupling. However, in Grand Unified Theories fermions in the same multiplet are expected to have 
a unified Yukawa coupling at the GUT scale and the fermion mass differences at low energies originate from radiative corrections. In all larger groups with the SM subgroups the b quark and the $\tau$-lepton are in the same multiplet and indeed, the ratio of their masses is in excellent agreement with the radiative
corrections  (including supersymmetric ones!)   between the GUT scale and the low scale, if one assumes $g_b =g_\tau$ at the GUT scale, see  Ref. \cite{deBoer:1994dg} and references therein. 

The branching ratios into the different final states are shown as function of the Higgs mass in Fig. \ref{f3}.
The $\tau$ final state is suppressed in comparison with the b quarks final state by the factor $m_\tau/3m_b$, where the factor 3 represents the colour factor.
At high masses the gauge boson final states dominate; the WW final state has a factor two higher branching fraction than the ZZ final state because the Feynman rules reduce the rate by a factor 1/n! for n identical particles in the final state.

Note that the neutrino stays massless within the SM, since no mass term
for the neutrino appears with the Higgs mechanism. Neutrinos can obtain a small mass via the see-saw mechanism, if they are Majorana particles and the right-handed neutrino exists with a high mass \cite{Mohapatra:1979ia,Yanagida:1980xy}.
%
\section{Evidence for the Higgs boson in CMS}
The LHC performance has been superb since initiating high energy pp collisions at
a centre-of-mass energy of 7 TeV in
early 2010, delivering 30 fb$^{-1}$
of data to CMS during the past three years, see Fig. \ref{f4} for details.  In 2012
the energy was raised to  4 TeV per beam  with an instantaneous luminosity
exceeding 7$\times$10$^{33}$ cm$^{-2}$s$^{-1}$. The average number of interactions per pp crossing (pile-up)
reached 21. The CMS detector was able to operate effectively in this high-occupancy environment with an efficiency above 90\%. 
The basic design of the general purpose CMS detector has been described in Ref. \cite{Chatrchyan:2008aa}
and extensive references about the performance of the detector for the Higgs channels has been given
in the long write-up of the Higgs discovery paper  \cite{Chatrchyan:2013lba}.
The most significant feature of the Compact Muon Solenoid (CMS) detector is the world's largest superconducting solenoid with a diameter of 6 m, 
a length of 13 m and a central field of 3.8 T. 
\begin{figure}[t]
\centering
\includegraphics[width=0.9\textwidth]{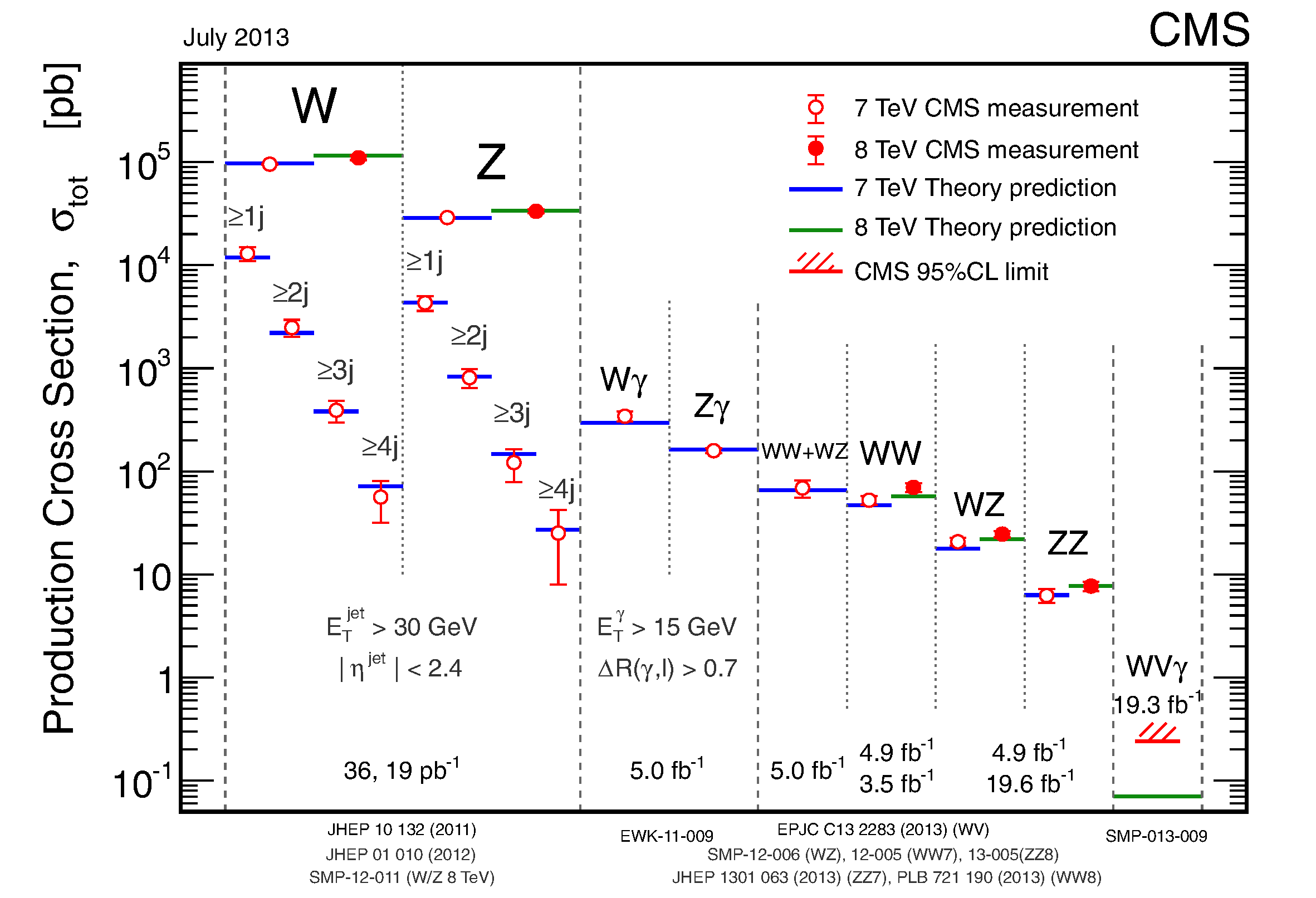}
\caption{ The vector boson production in the SM in comparison with CMS data. One observes that even the higher order QCD processes with multijets are well described. The total Higgs cross section for a 126 GeV Higgs is  in between the WZ and ZZ cross section.
}
\label{f6}
\end{figure}
\begin{figure}[t]
\centering
\includegraphics[width=0.8\textwidth]{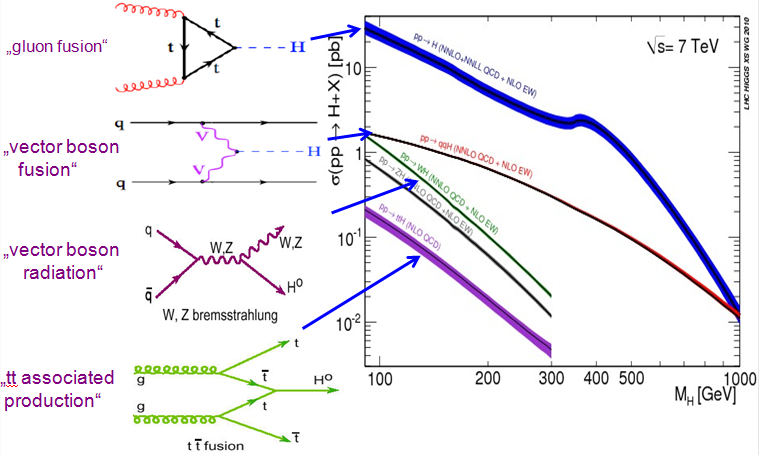}
\caption{ The  Higgs production cross sections in pp collisions at 7 TeV as function of the Higgs mass. The different contributions have been indicated. At 8 TeV the cross sections are typically 25-50\% higher.
}
\label{f7}
\end{figure}
The tracking is provided by the world's largest all-silicon tracker consisting of 200 m$^2$ of silicon sensors. 
The tracker is surrounded by a lead-tungstate scintillating crystal electromagnetic and brass/scintillator
 hadronic calorimeters, all located inside the magnet. 
 Outside the magnet is the tail-catcher of the hadronic calorimeter followed by the
instrumented iron return yoke, which serves as a multilayered muon
detection system.  Inside the yoke one has also a magnetic field, so the muon momentum is measured over a 
path length of almost 7 m, which provides an excellent invariant mass resolution. This is demonstrated in Fig. \ref{f5}, where all the resonances observed in $e^+e+-$ colliders are rediscovered by CMS in a much shorter time.
The CMS detector has extensive forward calorimetry, extending the pseudo-rapidity coverage
to $|\eta|  < 5.0$. 
Although the Higgs production cross section is small with respect to the SM processes, the superb performance of the LHC and the CMS and ATLAS detectors allowed for an amazingly speedy discovery of the Higgs boson. 
The SM vector boson production, one of the main backgrounds for the Higgs searches, is very well described by the CMS data, as shown in Fig. \ref{f6}. The Higgs cross section is of the same order as the WZ and ZZ production and is dominated by gluon-gluon fusion, as indicated in Fig. \ref{f7}.
\begin{figure}[]
\centering
\includegraphics[width=0.46\textwidth,height=0.35\textwidth]{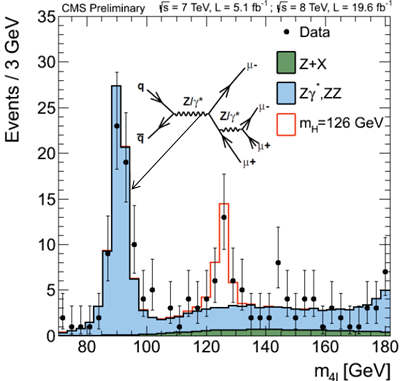}
\includegraphics[width=0.49\textwidth,height=0.37\textwidth]{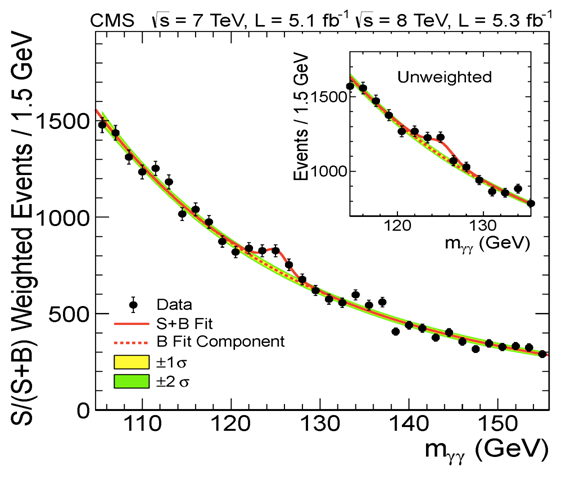}
\caption{ Invariant mass spectrum  of four leptons (left) and two photons (right). The enhancement at 126 GeV
is clearly visible in both cases.}
\label{f8}
\end{figure}
\begin{figure}[]
\centering
\includegraphics[width=0.6\textwidth]{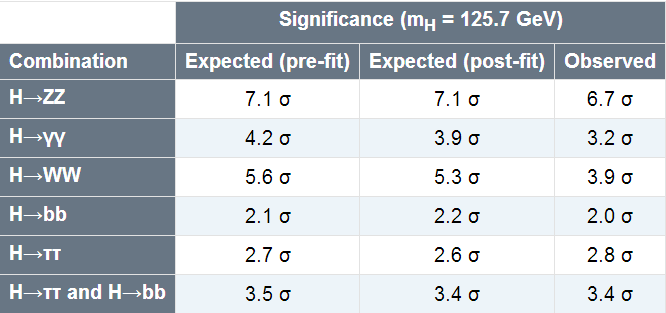}
\caption{ The median expected and observed significances of the excesses in the combination of
individual decay modes for a SM Higgs boson mass hypothesis of 125.7 GeV. The combined significance of
all channels is 9.4$\sigma$.}
\label{f9}
\end{figure}
\begin{figure}[]
\centering
\includegraphics[width=0.47\textwidth,height=0.35\textwidth]{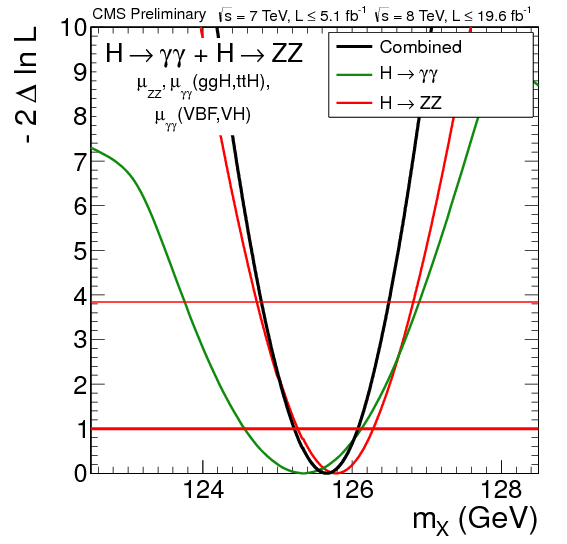}
\includegraphics[width=0.49\textwidth,height=0.35\textwidth]{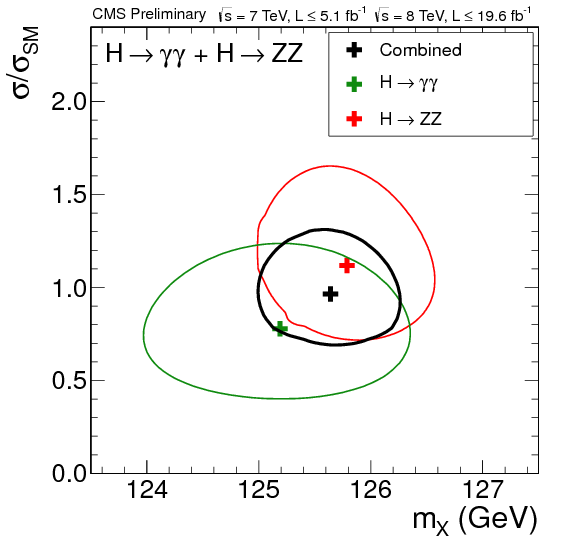}
\caption{  Left: $\chi^2$ distribution of a fit of the Higgs boson mass $m_H$ for the $\gamma\gamma$ and $4l$ final states separately and for their combination. The three indicated signal strengths have been left free in the fit together with all other nuisance parameters. 
Right: 2D 68\% CL contours for the  Higgs boson mass $m_H$ and the signal strength $\sigma/\sigma_{SM}$ for the $\gamma\gamma$ and $4l$ final states, and their combination. In this combination, the relative signal strengths are constrained by the expectations for the SM Higgs boson. From Ref. \cite{CMS:yva}.
}
\label{f10}
\end{figure}
\begin{figure}[]
\centering
\includegraphics[width=0.6\textwidth]{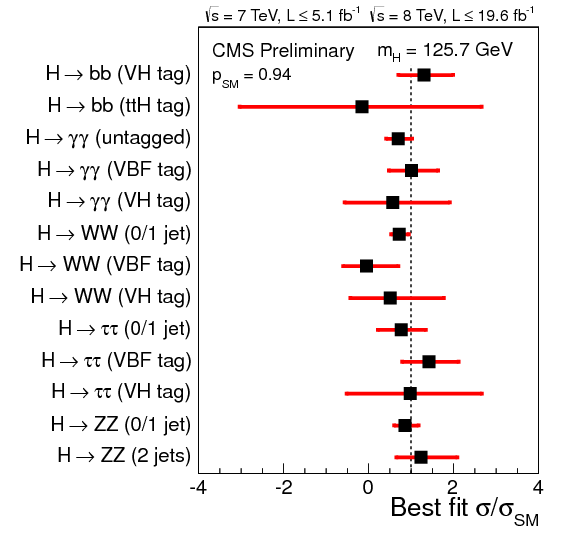}
\caption{ Measured signal strengths of the various channels in units of the SM signal strength.}
\label{f11}
\end{figure}
\begin{figure}[]
\centering
\includegraphics[width=0.45\textwidth,height=0.45\textwidth]{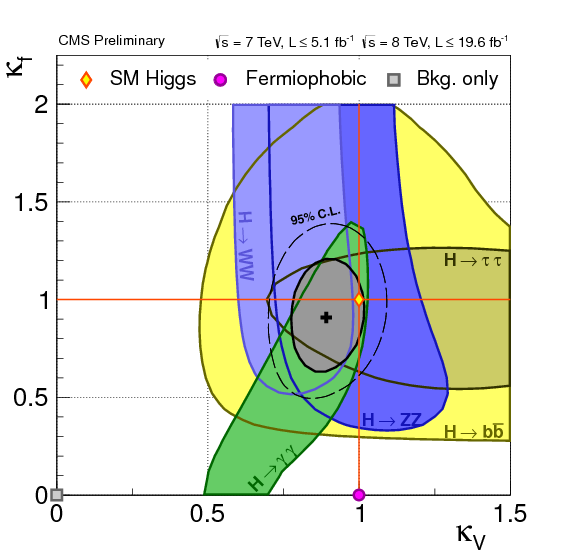}
\includegraphics[width=0.45\textwidth,height=0.43\textwidth]{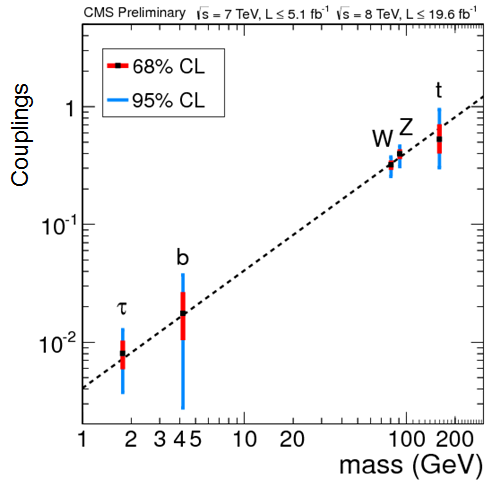}
\caption{ Left: 95\% C.L. contours in the plane of reduced vector boson coupling $\kappa_V$ and fermion couplings $\kappa_f$.  The data are compatible with the SM couplings, if $\kappa_f =\kappa_V =1$.
Right: Couplings to fermions and weak gauge bosons as function of mass. 
The errors and central values originate from the signal strengths of Fig. \ref{f12}. The following quark masses
have been used: $m_t~=~172.5$ GeV, $m_b(m_H=125.7$ GeV)~=~2.763 GeV. From Ref. \cite{CMS:yva}. 
}
\normalsize
\label{f12}
\end{figure}
The various Higgs decay channels investigated by CMS have been indicated 
by the circles on the vertical line at 126 GeV in Fig. \ref{f3}. 
 The $\gamma\gamma$ and $ZZ\rightarrow 4l$ final states have the highest significance, the reason being that in these final states a very good mass resolution becomes possible thanks to the absence of neutrinos and a small background from SM processes.
The invariant mass spectra of these final states are shown in Fig. \ref{f8}. The WW, bb and $\tau\tau$ channels have also been measured. A summary of the significances  is shown in Fig. \ref{f9}. The  ZZ channel has the highest mass resolution and leads to a mass of $m_h=125.8 ~\pm~ 0.5 (stat)~\pm ~0.2 (syst)$ GeV \cite{CMS:xwa}. The mass measurement from  the $\gamma\gamma$ channel is compatible with this most precise ZZ channel, as shown in Fig. \ref{f10}. As the statistical errors on the Higgs mass show, the $ZZ$ channel is
still statistically limited, so with more statistics in the future, one could hope for total errors of the order
of the systematic error of 0.2 GeV only. For the $\gamma\gamma$ channel the systematic error already dominates. 
The combined mass is \cite{CMS:yva}:  \bq m_h=125.7~\pm~0.3(stat.)~\pm~0.3(syst.)~{\rm GeV}.\eq

The signal strength of the various channels are all  compatible with the SM expectation, as shown in Fig. \ref{f11}.  One can quantify deviations from the SM couplings by  common scaling factors $\kappa_f$ and $\kappa_V$ for the coupling between the Higgs boson to fermions and vector bosons, respectively.
The allowed region in the $\kappa_f$-$\kappa_V$ plane is consistent with the SM expectation $\kappa_f =\kappa_V=1$, as shown in the left panel of Fig. \ref{f12}. The mass dependence of the couplings is demonstrated in  the right panel of Fig. \ref{f12}. The coupling  $g_{HVV}$ between the Higgs boson and the vector bosons is proportional to the mass squared, since it originates from the quadratic kinetic energy term in the Lagrangian. To get a linear relation between coupling and mass, the square root of the vector boson couplings was taken. The experimental value of these couplings
was derived from the signal strengths in Fig. \ref{f11}.
%
\begin{figure}[t]
\centering
\includegraphics[width=0.7\textwidth]{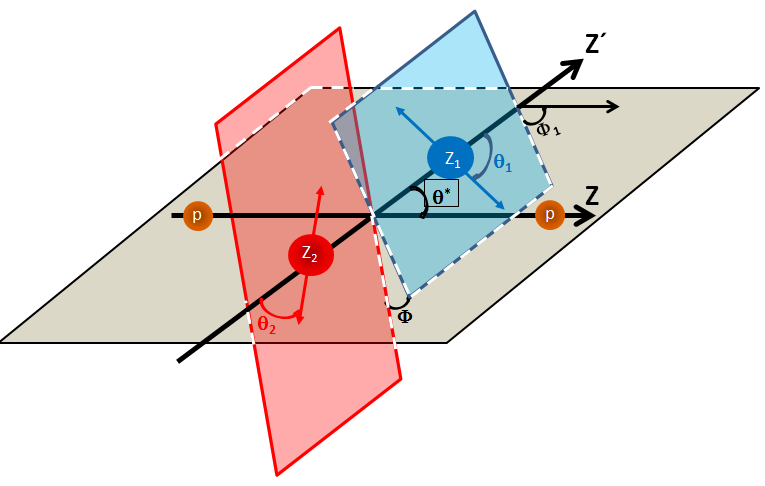}
\includegraphics[width=0.8\textwidth]{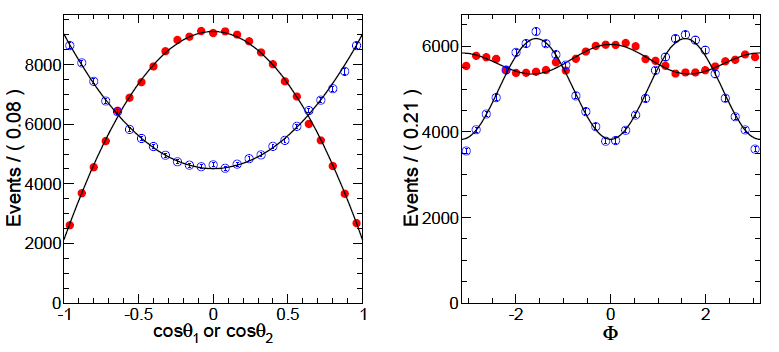}
\caption{Definition of decay angles (top panel) and sensitivity to the parity of the Higgs boson for positive (red, solid symbol) and negative parity (blue, open symbols)  for the most sensitive angles (bottom). Bottom plots from Ref. \cite{Gao:2010qx}.}
\label{f13}
\end{figure}
\begin{figure}[t]
\centering
\includegraphics[width=0.45\textwidth,height=0.45\textwidth]{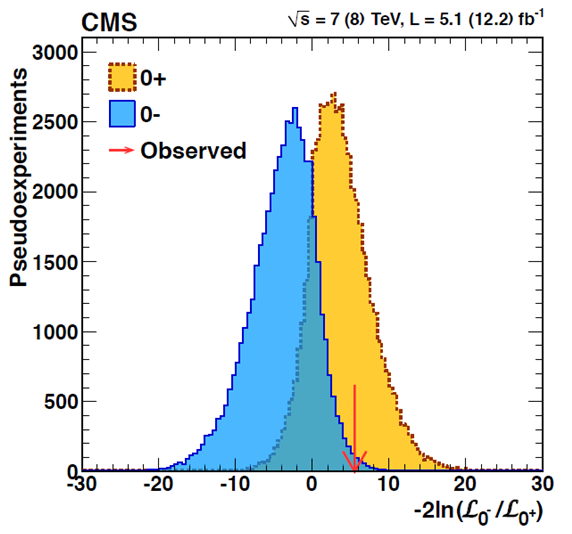}
\caption{ Probability density distribution of the $4l$ final state for a scalar and pseudo-scalar Higgs boson.
The red arrow shows the observed value of the statistic (likelihood ratio), which shows that the scalar boson is clearly preferred over the pseudo-scalar boson. The p-values are:  p(0$^+$) = 0.72 and p(0$^{–}$) = 0.072. The latter value corresponds to a probability of 2.4\%. From Ref. \cite{Chatrchyan:2012jja}.
}
\label{f14}
\end{figure}
\section{Is it Peter's Higgs?}
The most important properties of the Higgs boson are its mass, spin and parity. The mass is not predicted in the SM, only in its supersymmetric extension it is predicted to be below 130 GeV, as will be discussed below.
The mass has been discussed above. 
The spin 1 option is excluded by the virtue of the observation of the decay into two photons, since a particle with spin 1 does not couple to two identical particles with spin 1 
(Landau-Yang theorem, \cite{Landau:1948kw,Yang:1950rg}). This theorem also implies that a new
state with spin 1 can't be produced in gluon fusion, the dominant Higgs production channel.
Hence, only the spin 0 and spin $> 1$  options are options.
Here we concentrate on the most natural option of the lowest spin, especially since spin 2 and higher become model dependent. The models studied so far yield only a poor fit for spin 2 to the data \cite{Chatrchyan:2012jja}.

The spin 0 boson can be either a scalar (positive parity) or pseudo-scalar (negative parity).
Conservation of the angular momentum component along the decay axis fixes the spin projection of a spin 0 particle to $\lambda_h = \lambda_1 - \lambda_2=0$, where $\lambda_i$ are the helicities of the decay particles, e.g. the Z-bosons. 
 To study the parity one can look at the angular correlations in the decay products. The 5 angles in the 4-lepton final state defining completely the directions of the outgoing particles are depicted in Fig. \ref{f13}, top panel.
The explicit dependence on the angles has been given in Refs. \cite{Gao:2010qx,DeRujula:2010ys}. The largest sensitivity is in the helicity angles $\theta_i$ and $\Phi$, as shown in Fig. \ref{f13}, bottom panel.
The dependence on $\Phi$ illustrates the fact that
 in case of $0^+$ the polarization vectors of the outgoing bosons can only form a scalar product $\epsilon_1 \cdot \epsilon_2\propto \cos\Phi$, so the decay planes are preferentially aligned, while in case of $0^-$ the polarization vectors of the outgoing bosons can only form a pseudo-scalar product $\epsilon_1 \times \epsilon_2\propto \sin\Phi$,  in which case the decay planes are preferentially orthogonal.
 The result of the fit to the angular correlations clearly prefers the SM prediction $J^{P}=0^+$, as shown in
 Fig. \ref{f14}.

\section{Is the Higgs Boson the Origin of Mass in the Universe?}
As discussed above, the elementary particles obtain an effective mass from the interactions with the Higgs field.
However, the mass in the universe has little to do with the mass of elementary particles.
From cosmology we know that the energy of the universe has three main components \cite{Hinshaw:2012aka,Ade:2013zuv}:
\begin{itemize}
\item
5\% visible matter, but the mass is mainly provided by the mass of the nuclei (around 1 GeV/nucleon), not by the mass of the constituent quarks (around 1 MeV/quark for u and d quarks).
The main reason for the large difference in nucleon mass and quark mass is the binding energy of the quarks inside the nucleon, which is provided by the potential energy of the colour field, not by the Higgs field.
The quarks are kept inside the potential well of the colour field, in which  kinetic and potential energy are constantly exchanged. Since they are of the same order of magnitude, one can also picture the mass of nuclei
as the kinetic energy of its constituents, the quarks and gluons.
\item 25\% of the energy is provided by the dark matter. We do not know what is the nature of the dark matter, but we know that none of the known SM particles can describe its properties, so it is unlikely that the mass of the dark matter is provided by the interaction with the Higgs field. E.g. in supersymmetry the mass of the dark matter is provided by the breaking of this symmetry, not by the breaking of the electroweak symmetry.
\item
70\% of the energy is provided by the {dark energy}, a form of energy with negative pressure, which provides "repulsive" gravity to the matter in our universe. It was discovered by the accelerated expansion of the universe, for which Perlmutter, Schmidt and Riess got the Nobel prize in 2011 \cite{Riess:1998cb,Perlmutter:1998np}. 
Such a repulsive gravity was introduced by Einstein in his theory of general relativity as a cosmological constant in order to prevent a gravitational collapse of a static universe. Such a term is not needed in an expanding universe, but it is not forbidden either.
The most characteristic feature of the cosmological constant is its constant energy density in space, which naturally leads to "repulsive" gravity, see e.g. \cite{deBoer:1994dg}.
But this is also what the Higgs mechanism postulates: a Higgs field with a constant energy density everywhere in the vacuum ("vacuum energy"). From the Higgs potential one can easily estimate the energy density in the minimum with a nonzero vev to be 
(by inserting $\Phi^\dagger\Phi=v^2/2$ into Eq. \ref{potential} and replacing the potential parameters by the physical masses)\footnote{In SI units: $1~ GeV^4=(GeV/c^2 )(GeV^3/(hc)^3)= 10^{-24}~ g~ 10^{42}~ cm^{-3} = 10^{18} g/cm^3$, where we used one power of GeV to transfer to mass (via E=mc$^2$) and three powers to length via the de Broglie relation $E=pc=hc/\lambda$.}:
$V(\phi=\phi_0)=-m_H^2 m_W^2/2g^2\approx 10^8$ GeV$^4$ $\approx 10^{26}$ g/cm$^{3}.$
The total energy density  corresponds to the critical density for the universe, which is observed to be flat, i.e. the potential and kinetic energy cancel each other, so the total energy is zero: $\rho=2\cdot 10^{-29}$ g/cm$^3$.  This critical density is 55 orders smaller than the energy density in the Higgs field, estimated above. The smallness of the observed energy density in the universe is usually called the "cosmological constant problem", as first discussed by Weinberg \cite{Weinberg:1988cp,Weinberg:2000yb}.
The problem gets even more severe in Grand Unified Theories with Higgs fields near the Planck scale of $10^   {18}$ GeV, which leads to a 120 orders of magnitude problem between the Higgs energy density and the dark energy density. Although supersymmetry predicts Higgs particles at the GUT scale as well, the problem is here reduced, since  the contributions to the va\-cuum energy from fermions and bosons cancel. Since the SUSY particle masses are of the order of the TeV scale, i.e. at least an order of magnitude heavier than the SM particles, the cancellation is not perfect and the discrepancy is still 60 orders of magnitude. However, one can always shift the potential of the minimum of the Mexican hat to zero instead of putting zero at the center. Then there is no problem, except that one has to fine-tune this offset by 120 orders of magnitude to obtain the observed dark energy density. Hence, the "cosmological constant" problem or alternatively the question: "Why is the universe so empty?" is really an "order-of-magnitude" or "fine-tuning" problem.
\end{itemize}

\section{If the Higgs boson is not the Origin of Mass in the Universe, what is the Higgs Boson good for?}
As discussed above, most of the mass of the universe has nothing to do with the mass of elementary particles.
So one can ask: what is the Higgs boson good for? The answer is simple: without a Higgs boson our universe would be a completely different universe, in which life, as it is known to us, would not exist.
The simplest arguments are:
 \begin{itemize}
 \item
 if the electron would be massless, the atoms would not exist, since the Bohr radius of an atom is inversely proportional to the electron mass, so the atom would become infinite, i.e. no bound states would exist.
\item
without mass of the W- and Z-bosons the weak interactions would be not mass-suppressed at low energies. In this case fusion reactions in stars, like our sun, would be 5 orders of magnitude faster and our sun would not shine for $\approx 10^{10}$  years, but only  $\approx 10^5$ years, i.e. too short for live to develop on our planet.
\end{itemize}

\section{What is so special about the observed Higgs Boson?}
In the SM the Higgs mass is not predicted and can be anywhere between the electroweak and TeV scale. 
It can't be much heavier, since else the scattering of  two longitudinally polarized W-bosons would start to violate unitarity at energies above 1 TeV. The exchange of a scalar particle  compensates the 
 exchange of gauge bosons in the Feynman diagrams. 
In the supersymmetric extension of the SM the Higgs mechanism does not need to be introduced ad hoc, but is predicted by radiative corrections: the Higgs mass parameters in the potential can be positive at the GUT scale, but get negative radiative corrections from top and bottom quarks.
Such radiative corrections can drive the mass parameter negative, but this is exactly what is needed to break the symmetry, i.e. change a parabolic potential to a Mexican hat potential in case of a single Higgs doublet. In order to obtain symmetry breaking at the electroweak scale the radiative corrections have to be strong enough to drive a positive mass term at the GUT scale negative at the electroweak scale. This is only possible for top masses between 140 and 200 GeV (see e.g. \cite{deBoer:1994dg}). 
This prediction of EWSB is a strong argument in favour of supersymmetry, but what is even stronger:
supersymmetry predicts the lightest Higgs boson to have a mass below 130 GeV and it is observed to be below 130 GeV! This may be the strongest hint for SUSY  in spite of the fact that no SUSY particles have been observed so far. In the next section we detail a little more the impact of the observation of a 126 GeV Higgs on Supersymmetry.
\section{Expectations from Supersymmetry}
As mentioned above, supersymmetry predicts a Higgs mass below 130 GeV, as observed.
However, also a few problems exist. First of all, at tree level the Higgs boson mass is predicted to be 
below the $Z^0$-mass in the Minimal Supersymmetric Standard Model (MSSM)  and to obtain a mass of 126 GeV requires large radiative corrections.
 The dominant radiative corrections come 
 from  loops with the third generation quarks and their supersymmetric partners, mainly the stop squarks $\tilde{t}$, which can be written as (see Ref. \cite{Djouadi:2005gj} and references therein):
\bq
\Delta {m}_{h}^2 \sim   \frac{3\, \bar{m}_t^4}{2\pi^2 v^2\sin^
2\beta} \left[ \log \frac{M_S^2}{\bar{m}_t^2} + \frac{X_t^2}{2\,M_S^2} \left( 1 -
\frac{X_t^2}{6\,M_S^2} \right) \right]
\label{higgscorr}
\eq
where $M_S$ is the arithmetic  average of the stop masses
$M_S =\frac{1}{2} (m_{\tilde{t}_1}+m_{\tilde{t}_2})$,
$X_{t}$ is the stop mixing parameter, and
$\bar{m}_t$ is the running ${\rm \overline{MS}}$ top quark mass to account
for the leading two--loop QCD and electroweak corrections in a RG improvement.
 The corrections become maximal, if the stop mixing parameter  $X_t=\sqrt{6} M_S$. However, this so-called maximal mixing scenario is not allowed, if the parameters are defined at a high scale,
since  the quasi-fixed-point solutions of the renormalization group equations forbid maximal mixing in the stop sector, in which case  the stop masses typically need to be 4 TeV or above for a 126 GeV Higgs (see e.g.   \cite{Beskidt:2013gia}).
 
Alternatively, one can consider the Next-to-Minimal Supersymmetric Standard Model (NMSSM), which has,
 in addition to the two Higgs doublets of the MSSM, an additional Higgs singlet. 
  Details on the NMSSM Higgs sector can be found in Refs. \cite{Miller:2003ay,Ellwanger:2009dp}.
 The NMSSM has attracted much attention in the last year, 
since the  additional contributions at tree level from the mixing with the singlet avoids  
 the need  for multi-TeV  stop quarks to reach a Higgs mass of 126 GeV.
 A Higgs singlet had been proposed before the Higgs boson discovery for various reasons, among them that it solves the so-called $\mu$-problem: 
 electroweak symmetry breaking only works, if the Higgs mixing parameter $\mu$ between 
 the two Higgs doublets is of the order of the electroweak scale, although this superpotential parameter 
 with a dimension of mass  a priori could take any value up to the GUT scale. 
 In the NMSSM the $\mu$ parameter originates from the vev of the singlet Higgs boson, 
 which naturally takes  a value of the electroweak scale. 
 This has very specific consequences for the Higgs sector, especially it can lead to double Higgs production and
 invisible Higgs decays. For a detailed discussion of these interesting aspects, see Ref. \cite{Beskidt:2013gia} and references therein. Here we only summarize the salient features.
\begin{figure}[]
\centering
\includegraphics[width=0.49\textwidth,height=0.4\textwidth]{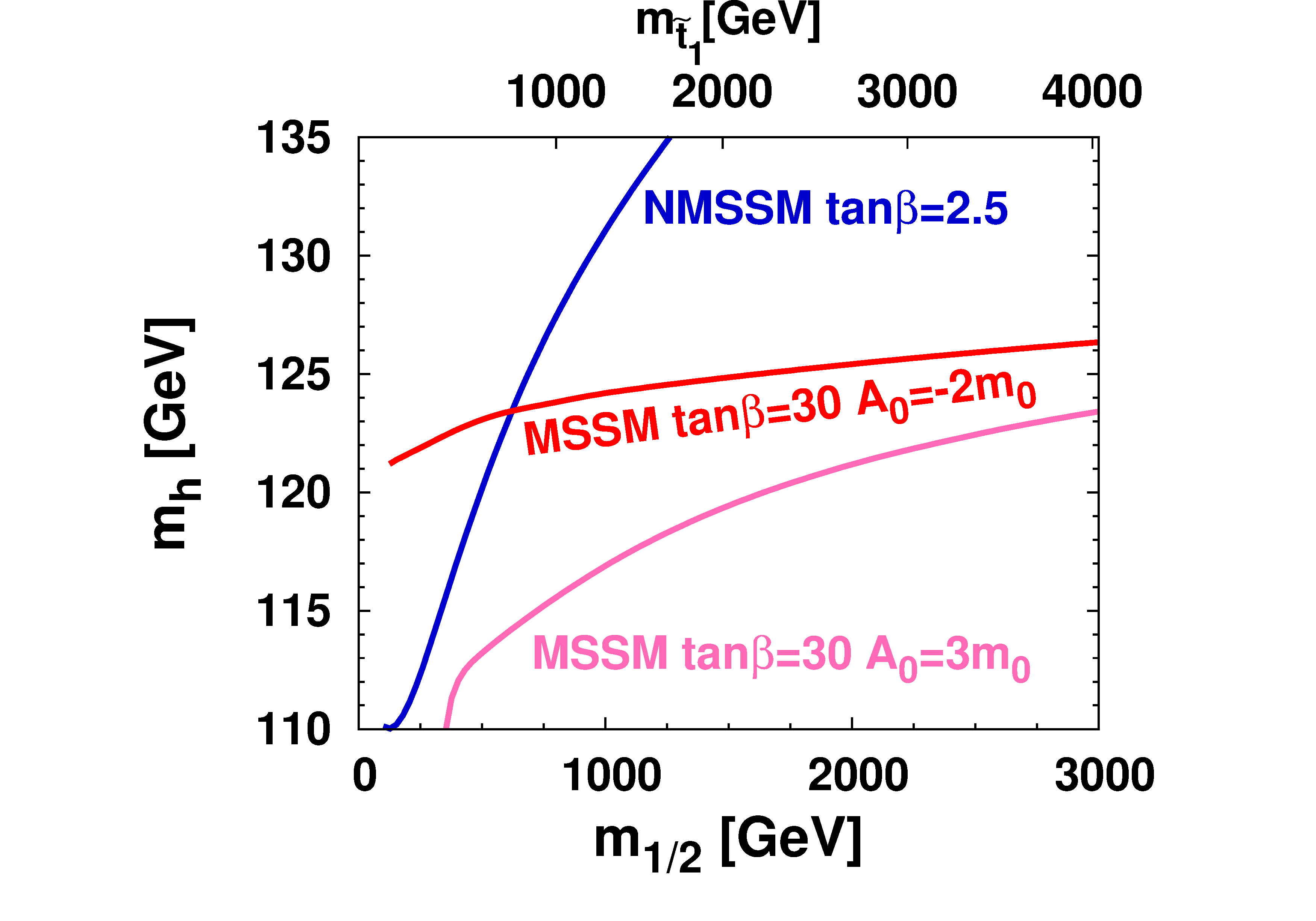}
\includegraphics[width=0.49\textwidth,height=0.4\textwidth]{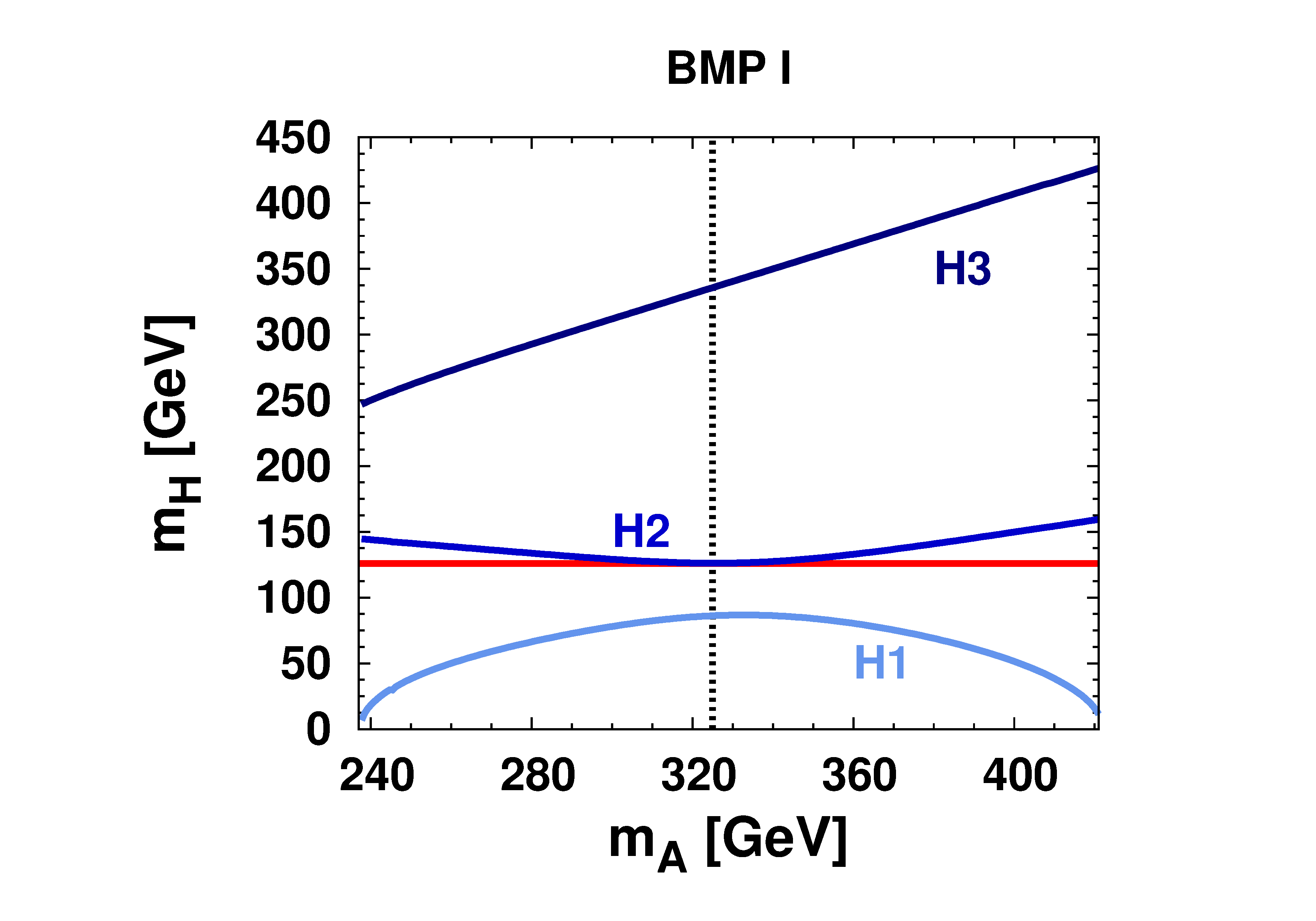}
\caption{ Left: comparison of the Higgs boson mass within the MSSM and NMSSM plotted as function of 
the GUT scale mass parameter $\mhalf$ for fixed values of all other parameters. 
The stop mass is indicated at the top. Clearly, in the NMSSM a 126 GeV Higgs mass can be reached
for lower stop masses. 
Right: The three Higgs scalar boson masses of the NMSSM as function of the pseudo-scalar Higgs mass $M_A$. 
The horizontal (red) line corresponds to a Higgs boson mass of 126 GeV. 
The vertical line shows the value of $M_A$ which corresponds to a 126 GeV Higgs with SM couplings. 
From Ref. \cite{Beskidt:2013gia}.}
\label{f15}
\end{figure}
In supersymmetry the Higgs sector has two complex Higgs doublets with 8 degrees of freedom, 5 degrees of freedom are left after the W$^+$, W$^-$ and Z$^0$ bosons obtain mass, so 5 Higgs bosons are predicted in the minimal supersymmetric standard model (MSSM). 
In the NMSSM  one has two degrees of freedom more because of the complex singlet, so one expects 7 Higgs bosons instead of 5. They mix and the lightest Higgs mass gets already a contribution at tree level from this mixing, so a mass of 126 GeV can be reached for  lower stop masses,
as demonstrated in Fig. \ref{f15}, left panel. The two lightest scalar Higgs bosons are close in mass, so they have the strongest mixing, which in turn depends on the value of the  nearly degenerate and decoupled heavier Higgs masses, as demonstrated in the right panel of Fig. \ref{f15}. Minimal mixing occurs at the maximum of the $H_1$ mass and for different values of the pseudo-scalar $M_A$ mass mixing increases, thus lowering one eigenvalue and increasing the other one. Since the lightest Higgs boson has a large singlet component, which does not couple to SM particles, one can easily obtain for $H_2$ couplings deviating from the SM couplings by the mixing with $H_1$. For the value indicated by the dotted vertical line $H_2$ has SM couplings. The value of $M_A$ is  proportional to the $\mu$ parameter, which is the vev of the singlet, so of the order of the weak scale.
  So the heavier Higgs bosons are expected to be naturally light, although by tuning the trilinear couplings to large values, TeV scale heavy boson masses can be obtained. The lowest value of the degenerate heavier Higgs boson is around 250 GeV, since the value of the $\mu$ parameter determines also the lightest chargino mass, for which the lower limit is 104 GeV from the LEP searches \cite{Beringer:1900zz}.
  There is also a lower limit on the lightest Higgs $H_1$ from the relic density: if it is below 60 GeV it would not generate the correct relic density. Since it is the lightest Higgs it cannot have a mass above 126 GeV, which is typically the SM-like Higgs $H_2$ \footnote{Solutions, where $H_1$ is SM-like are also possible, but then $H_2$ is usually very close or even degenerate with $H_2$, but these solutions have a rather restricted region of parameter space.}.
    
    The mixing terms lead also to quite different decay channels between the
MSSM and NMSSM, especially  the decay of the heavier scalar Higgs boson into two lighter Higgses ($H_3\rightarrow H_2+H_1$) becomes possible and is as large as 41\% for the benchmark point depicted in Fig. \ref{f15}. This leads to double Higgs production, i.e. one observes two Higgs bosons in a single event, of which  one has a mass of 126 GeV. Double Higgs production is practically absent in the MSSM, so if observed, this could be a unique signature for the NMSSM \cite{Kang:2013rj}. The additional singlet leads also to an additional singlino, which mixes with the other neutralinos and the lightest neutralino is usually singlino-like. The lightest neutralino is usually the lightest supersymmetric particle (LSP), which is stable, if  R-parity is conserved. Its self-couplings turn out to yield an annihilation cross section in excellent agreement with the observed relic density, so it is a perfect dark matter candidate.
Also the couplings to nuclei are well below the limits set by the direct dark matter searches because of its large singlino-component. Given this  Higgs\-ino character of the LSP, the heavier Higgses tend to have a significant branching ratio into LSP's by the strong couplings between the Higgs bosons in the NMSSM , which would lead to Higgs events with missing energy and invisible Higgs decays.
So it will be very interesting to search for additional Higgs bosons and observe their decay modes. 
\begin{figure}[]
\centering
\includegraphics[width=0.75\textwidth,angle=0,height=0.65\textwidth]{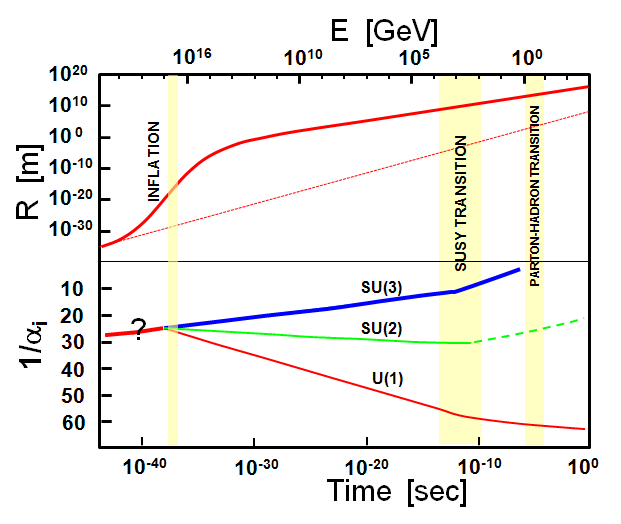}
\caption{Evolution of the scaling factor of the universe (top) and evolution of the gauge couplings (bottom) 
as function of time after the Big Bang. From Ref. \cite{deBoer:1994dg}.}
\label{f16}
\end{figure}
\section{Summary}
Observing the Higgs boson roughly 50 years after its prediction is a great triumph for particle physics, both from the theory side and the experimental side, which required about 20 years of planning and building the LHC accelerator and its large detectors with collaborations comprising several thousands of physicists.

The most interesting question for the future will be: are there more Higgs bosons? This is expected in two Higgs doublet models, like Supersymmetry. The fact that the observed Higgs mass is within the range predicted by supersymmetry pins the hope on more Higgs bosons. If Supersymmetry is indeed confirmed in the future, it would 
 be an even greater triumph for particle physics, since it might provide a candidate for the elusive dark matter. Furthermore, it might the perfect candidate for a Grand Unified Theory, given the fact that   in Supersymmetry the gauge couplings unify for SUSY masses at the TeV scale \cite{Amaldi:1991cn}. 
A Grand Unified Theory would imply several phase transitions during the evolution of the universe from the Planck
temperature of $10^{32}$ K to the 2.7 K observed today, which are needed to solve many of the questions 
posed by cosmology.
Among them: the baryon asymmetry in our universe and
inflation, which is the only viable solution to explain the
horizon problem,
 the flatness problem, the
magnetic monopole problem, and
the smoothness problem, see e.g. Ref. \cite{deBoer:1994dg}.
 In this case the evolution of the universe could indeed be, as depicted in Fig. \ref{f16}: the early universe was governed by a unified force, which breaks into the well known forces of the SM, described by the $SU(3)\otimes SU(2) \otimes U(1)$ groups. This symmetry breaking might be the source of a dominance of vacuum energy at $t\approx 10^{-37}$ s after the big bang, leading to the burst of inflation, needed to obtain a flat and isotropic universe. In the time between $t\approx 10^{-37}$
 and about 1 ps the universe cools from a temperature corresponding to  $10^{16}$ to $10^{3}$ GeV and nothing happens (the Great Desert).
  Near the TeV scale   the SUSY  particles decouple and at a scale of 100 GeV electroweak symmetry breaking occurs, leading to a decoupling of the weak force.
 This picture was devised more than 20 years ago \cite{deBoer:1994dg}, but it has become more plausible after we know that spontaneous symmetry breaking can be caused by a scalar boson and  the mass of the newly found  Higgs boson is within  the narrow range predicted by this scenario.





\footnotesize
\providecommand{\href}[2]{#2}\begingroup\raggedright\endgroup

\normalsize
\end{document}

%% file: config.tex
\newlength{\dslashwidth}

\newcommand{\LL}{{\ifmmode {\cal L} \else ${\cal L}$\fi}}

\newcommand{\beq}{\begin{equation}} 
\newcommand{\eeq}{\end{equation}}
\newcommand{\beqa}{\begin{eqnarray}} 
\newcommand{\eeqa}{\end{eqnarray}}
\newcommand{\newc}{\newcommand}
\newcommand{\bq}{\begin{equation}}
\newcommand{\eq}{\end{equation}}
\newcommand{\ba}{\begin{array}}
\newcommand{\ea}{\end{array}}
\newcommand{\bqa}{\begin{eqnarray}}
\newcommand{\eqa}{\end{eqnarray}}

\newcommand{\ra}{\rightarrow}
\newcommand{\lnf}{{\ifmmode \Lambda^{(N_f)} \else $\Lambda^{(N_f)}$\fi}}
\newcommand{\ms}{{\ifmmode \overline{MS} \else $\overline{MS}$\fi}}
\newcommand{\dr}{{\ifmmode \overline{DR} \else $\overline{DR}$\fi}}
\newcommand{\lms}{{\ifmmode \Lambda^{(5)}_{\overline{MS}} \else $\Lambda^{(5)}_{\overline{MS}}$\fi}}
\newcommand{\lam}{{\ifmmode \Lambda \else $\Lambda$\fi}}
\newcommand{\mev}{{\ifmmode {\rm MeV} \else ${\rm MeV}$\fi}}
\newcommand{\gev}{{\ifmmode {\rm GeV} \else ${\rm GeV}$\fi}}
\newcommand{\gevc}{{\ifmmode {\rm GeV/c^2} \else ${\rm GeV/c^2}$\fi}}
\newcommand{\tev}{{\ifmmode {\rm TeV} \else ${\rm TeV}$\fi}}
\newcommand{\tevc}{{\ifmmode {\rm TeV/c^2} \else ${\rm TeV/c^2}$\fi}}
\newcommand{\lp}{{\ifmmode L^+  \else $L^+$\fi}}
\newcommand{\lm}{{\ifmmode L^-  \else $L^-$\fi}}
\newcommand{\mlp}{{\ifmmode M(L^-) \else $M(L^-)$\fi}}
\newcommand{\mlz}{{\ifmmode M(L^0) \else $M(L^0)$\fi}}
\newcommand{\lz}{{\ifmmode L^0 \else $L^0$\fi}}
\newcommand{\ev}{{\ifmmode GeV/c^2 \else $GeV/c^2$\fi}}
\newcommand{\tri}{{\ifmmode \triangleup \else $\triangleup$\fi}}
\newcommand{\unl}{{\ifmmode U_{lL^0} \else $U_{lL^0}$\fi}}\newcommand{\gL}{{\ifmmode g_L \else $g_{L}$\fi}}
\newcommand{\gR}{{\ifmmode g_R  \else $g_{R}$\fi}}
\newcommand{\gumu}{{\ifmmode \gamma^{\mu} \else $\gamma^{\mu}$\fi}}
\newcommand{\gunu}{{\ifmmode \gamma^{\nu} \else $\gamma^{\nu}$\fi}}
\newcommand{\gdmu}{{\ifmmode \gamma_{\mu} \else $\gamma_{\mu}$\fi}}
\newcommand{\gdnu}{{\ifmmode \gamma_{\nu} \else $\gamma_{\nu}$\fi}}
\newcommand{\stw}{{\ifmmode\sin^2\theta_W \else $\sin^{2}\theta_{W}$ \fi}}
\newcommand{\sws}{{\ifmmode \;\sin^2\theta_W  \else $\;\sin^{2}\theta_{W}$ \fi}}
\newcommand{\cws}{{\ifmmode \;\cos^2\theta_W  \else $\;\cos^{2}\theta_{W}$ \fi}}
\newcommand{\sw}{{\ifmmode \;\sin\theta_W  \else $\sin\theta_{W}$ \fi}}
\newcommand{\cw}{{\ifmmode \;\cos\theta_W  \else $\;\cos\theta_{W}$ \fi}}
\newcommand{\tw}{{\ifmmode \;\tan\theta_W  \else $\;\tan\theta_{W}$ \fi}}
\newcommand{\qq}{{\ifmmode q\overline{q} \else $q\overline{q}$\fi}}
\newcommand{\lR}{{\ifmmode l_R \else $l_R$\fi}}
\newcommand{\lL}{{\ifmmode l_L \else $l_L$\fi}}
\newcommand{\nt}{{\ifmmode \nu_{\tau} \else $\nu_{\tau}$\fi}}
\newcommand{\nuR}{{\ifmmode \nu_R  \else $\nu_R$\fi}}
\newcommand{\nuL}{{\ifmmode \nu_L  \else $\nu_L$\fi}}
\newcommand{\qR}{{\ifmmode g_R  \else $q_R$\fi}}
\newcommand{\qL}{{\ifmmode q_L  \else $q_L$\fi}}
\newcommand{\qRp}{{\ifmmode q_R'  \else $q_{R}$'\fi}}
\newcommand{\qLp}{{\ifmmode q_L'  \else $q_{L}$'\fi}}
\newcommand{\est}{{\ifmmode e^{\bf \ast} \else $e^{\bf \ast}$\fi}}
\newcommand{\lst}{{\ifmmode l^{\bf \ast} \else $l^{\bf \ast}$\fi}}
\newcommand{\must}{{\ifmmode \mu^{\bf \ast} \else $\mu^{\bf \ast}$\fi}}
\newcommand{\taust}{{\ifmmode \tau^{\bf \ast} \else $\tau^{\bf \ast}$ \fi}}
\newcommand{\pperp}{{\ifmmode p_t  \else $p_t$\fi}}
\newcommand{\et}{{\ifmmode E_t  \else $E_t$\fi}}
\newcommand{\xt}{{\ifmmode x_t  \else $x_t$\fi}}
\newcommand{\smumu}{{\ifmmode \sigma_{\mu\mu}  \else $\sigma_{\mu\mu}$ \fi}}
\newcommand{\eg}{{\ifmmode e\gamma  \else $e\gamma$\fi}}
\newcommand{\epem}{{\ifmmode e^+e^-  \else $e^+e^-$\fi}}
\newcommand{\lplm}{{\ifmmode L^+L^-  \else $L^+L^-$\fi}}
\newcommand{\pp}{{\ifmmode p\overline p  \else $p\overline p$\fi}}
\newcommand{\llz}{{\ifmmode L^0\overline{L}^0 \else $L^0\overline{L}^0$\fi}}
\newcommand{\epemt}{{\ifmmode e^+e^- \to  \else $e^+e^- \to$\fi}}
\newcommand{\eb}{{\ifmmode E_{beam}  \else $E_{beam}$\fi}}
\newcommand{\ip}{{\ifmmode pb^{-1}  \else $pb^{-1}$\fi}}
\newcommand{\upm}{{\ifmmode ^{\pm}  \else $^{\pm}$\fi}}
\newcommand{\de}{{\ifmmode ^{\circ}  \else $^{\circ}$ \fi}}
\newcommand{\appr}{{\ifmmode \sim \else $\sim$ \fi}}
\newcommand{\corresp}{{\ifmmode \stackrel{\wedge}{=} \else $\stackrel{\wedge}{=}$ \fi}}
\newcommand{\sqrts}{{\ifmmode \sqrt{s} \else $\sqrt{s}$\fi}}
\newcommand{\zz}{{\ifmmode Z^0  \else $Z^0$\fi}}
\newcommand{\mz}{{\ifmmode M_{Z}  \else $M_{Z}$\fi}}
\newcommand{\mzs}{{\ifmmode M_{Z}^2  \else $M_{Z}^2$\fi}}
\newcommand{\mw}{{\ifmmode M_{W}  \else $M_{W}$\fi}}
\newcommand{\mws}{{\ifmmode M_{W}^2  \else $M_{W}^2$\fi}}
\newcommand{\mh}{{\ifmmode M_{Higgs}  \else $M_{Higgs}$\fi}}
\newcommand{\gt}{{\ifmmode \Gamma_{tot} \else $\Gamma_{tot}$\fi}}
\newcommand{\msusy}{{\ifmmode M_{SUSY}  \else $M_{SUSY}$\fi}}
\newcommand{\msusys}{{\ifmmode M_{SUSY}^2  \else $M_{SUSY}^2$\fi}}
\newcommand{\su}{{\ifmmode SU(3)_C\otimes\- SU(2)_L\otimes\- U(1)_Y \else $SU(3)_C\otimes SU(2)_L\otimes U(1)_Y$\fi}}
\newcommand{\suthree}{{\ifmmode SU(3)_C  \else $SU(3)_C$\fi}}
\newcommand{\sutwo}{{\ifmmode  SU(2)_L\otimes U(1)_Y \else $SU(2)_L\otimes U(1)_Y$\fi}}
\newcommand{\taup}{{\ifmmode \tau_{proton} \else $\tau_{proton}$\fi}}
\newcommand{\as}{{\ifmmode \alpha_{s}  \else $\alpha_{s}$\fi}}
\newcommand{\mgut}{{\ifmmode M_{GUT}  \else $M_{GUT}$\fi}}
\newcommand{\mguts}{{\ifmmode M_{GUT}^2  \else $M_{GUT}^2$\fi}}
\newcommand{\mzero}{{\ifmmode m_0        \else $m_0$\fi}}
\newcommand{\mhalf}{{\ifmmode m_{1/2}    \else $m_{1/2}$\fi}}
\newcommand{\sq}{{\ifmmode \tilde{q}    \else $\tilde{q}$\fi}}
\newcommand{\gl}{{\ifmmode \tilde{g}    \else $\tilde{g}$\fi}}
\newcommand{\mb}{{\ifmmode m_{b}    \else $m_{b}$\fi}}
\newcommand{\mt}{{\ifmmode m_{t}    \else $m_{t}$\fi}}
\newcommand{\mts}{{\ifmmode m_{t}^2    \else $m_{t}^2$\fi}}

\newcommand{\mtau}{{\ifmmode m_{\tau}  \else $m_{\tau}$\fi}}
\newcommand{\dpp}{{\ifmmode \delta_{pert} \else $\delta_{pert}$\fi}}
\newcommand{\dnp}{{\ifmmode\delta_{non-pert}\else$\delta_{non-pert}$\fi}}
\newcommand{\dew}{{\ifmmode \delta_{\rm EW}\else $\delta_{\rm EW}$\fi}}
\newcommand{\rt}{{\ifmmode R_{\tau}  \else $R_{\tau} $\fi}}
\newcommand{\rz}{{\ifmmode R_{Z}  \else $R_{Z} $\fi}}

\newcommand{\swb}{{\ifmmode \sin^2\theta_{\overline{MS}} \else $\sin^2\theta_{\overline{MS}}$\fi}}
\newcommand{\cwb}{{\ifmmode \cos^2\theta_{\overline{MS}} \else $\cos^2\theta_{\overline{MS}}$\fi}}

\newc\AIPCP[3] {{\em AIP Conf. Proc.} {\bf #1} (#2) #3}
\newc\AJ[3] {{\em Astrophys. J.} {\bf #1} (#2) #3}
\newc\AMS[3] {{\em Ann. Math. Statist.} {\bf #1} (#2) #3}
\newc\AP[3] {{\em Ann. Phys.} {\bf #1} (#2) #3}
\newc\APJ[3] {{\em Astropart. J.} {\bf #1} (#2) #3}
\newc\APP[3] {{\em Astropart. Phys.} {\bf #1} (#2) #3}
\newc\APS[3] {{\em Astrophys. J. Suppl.} {\bf #1} (#2) #3}
\newc\ARNPS[3] {{\em Ann. Rev. Nucl. Part. Sci.} {\bf C#1} (#2) #3}
\newc\BA[3] {{\em Bayesian Anal.} {\bf C#1} (#2) #3}
\newc\CPC[3] {{\em Comput. Phys. Commun.} {\bf C#1} (#2) #3}
\newc\CP[3] {{\em Contemp. Phys.} {\bf #1} (#2) #3}
\newc\EPJ[3] {{\em Euro. Phys. Journ.} {\bf C#1} (#2) #3}
\newc\JCAP[3] {{\em JCAP} {\bf #1} (#2) #3}
\newc\JHEP[3] {{\em JHEP} {\bf #1} (#2) #3}
\newc\JPG[3] {{\em J. Phys.} {\bf G #1} (#2) #3}
\newc\IJMP[3] {{\em Int. J. Mod. Phys.} {\bf A #1} (#2) #3}
\newc\MNRAS[3] {{\em Mon. Not. Roy. Astron. Soc.} {\bf #1} (#2) #3}
\newc\MPL[3] {{\em Mod. Phys. Lett.} {\bf A #1} (#2) #3}
\newc\NAR[3] {{\em New Astron. Rev.} {\bf #1} (#2) #3}
\newc\NCA[3] {{\em Nuovo Cimento} {\bf #1} (#2) #3}
\newc\NIM[3] {{\em Nucl. Instrum. Methods} {\bf #1} (#2) #3}
\newc\NIMA[3] {{\em Nucl. Instrum. Methods} {\bf A #1} (#2) #3}
\newc\NAT[3] {{\em Nature} {\bf #1} (#2) #3}
\newc\NPB[3] {{\em Nucl. Phys.} {\bf B #1} (#2) #3}
\newc\NPA[3] {{\em Nucl. Phys.} {\bf A #1} (#2) #3}
\newc\NPPS[3] {{\em Nucl. Phys. Proc. Suppl.} {\bf #1} (#2) #3}
\newc\PLB[3] {{\em Phys. Lett.} {\bf B #1} (#2) #3}
\newc\PR[3] {{\em Phys. Rep.} {\bf #1} (#2) #3}
\newc\PRL[3] {{\em Phys. Rev. Lett.} {\bf #1} (#2) #3}
\newc\PRD[3] {{\em Phys. Rev.} {\bf D #1} (#2) #3}
\newc\PRC[3] {{\em Phys. Rev.} {\bf C #1} (#2) #3}
\newc\PTP[3] {{\em Prog. Theor. Phys.} {\bf #1} (#2) #3}
\newc\RMP[3] {{\em Rev. Mod. Phys.} {\bf #1} (#2) #3 }
\newc\RPP[3] {{\em Rept. Prog. Phys.} {\bf #1} (#2) #3 }
\newc\SC[3] {{\em Science} {\bf #1} (#2) #3 }
\newc\ZPC[3] {{\em Z. Phys.} {\bf C #1} (#2) #3}
\newc\Err[3] {{\em Erratum-ibid.} {\bf #1} (#2) #3 }